\documentclass[11pt]{nih2}

\usepackage[pdftex]{graphicx}
\usepackage{fancyhdr}
\begin{document}
\parskip=8pt
\parindent=0pt
\noindent
\begin{center}
\LARGE
\textbf{Biophysics-based critique of \\ the assisted discharge mechanism hypothesis}
\normalsize

\vspace{0.09in}

Julie V. Stern$^1$,
Thiruvallur R. Gowrishankar$^1$,
Kyle C. Smith$^{1}$,
and
James C. Weaver$^{1,*}$,

$^1$Harvard-MIT Division of Health Sciences and Technology,
Institute for Medical Engineering and Science,
Massachusetts Institute of Technology, Cambridge, MA, USA;

$^*$Corresponding author

\vspace{0.12in}
\end{center}

\vspace{0.23in}
\normalsize

%
\textbf{
\noindent
Cell experiments with large, short electric field pulses of opposite polarity reveal a remarkable
phenomenon:  Bipolar cancellation (BPC).  Typical defining experiments involve quantitative 
observation of tracer molecule influx at times of order 100 s post pulsing.  Gowrishankar et al.
BBRC 2018 503:1194-1199 shows that long-lived pores and altered partitioning or hindrance due to inserted
occluding molecules can account for BPC.  In stark contrast, the Assisted Discharge (AD) hypothesis,
Pakhomov et al. CellMolLifeSci 2014 71(22):4431-4441; Fig. 6, only involves early times of a microsecond down to
nanoseconds.  Further, well established terminology for cell membrane discharge relates to 
membrane potential decays shortly after pulsing.  Discharge is silent on molecular or ionic 
transport, and does not address the fact that tracer molecule uptake vs time is measure at
about 100s after pulsing ceases.  Our critique of AD notes that
there can be an association of AD with BPC, but associations are only necessary, not sufficient.
A BPC mechanism hypothesis must be shown to be causal, able to describe time-dependent molecular
influx.  The two hypotheses involve very different time-scales (less than a microsecond vs 100 s)
and very different quantities (volts/s vs molecules/s).  Unlike pore-based hypotheses the
AD hypothesis lacks explicit molecular transport mechanisms, and does not address the greatly
delayed measured molecular uptake.  We conclude that AD is an implausible candidate for explaining
BPC.  
}

%

\ \\
\noindent
\textbf{Basic approach:} Use general features of established science for 
plausibility estimates.  This biophysics/physics method has often been 
used to obtain rough estimates of the plausibility of reported results, 
new concepts, theoretical constructs, etc. Testing
is widely used, and can be extended to biophysics, including 
bioelectrics. In the present (biophysics of bioelectrics) order of 
magnitude (OOM) estimates can be compelling.

\noindent
\textbf{Four examples of order of magnitude estimates of plausibility using 
generally accepted science.} These are:
\begin{itemize}
\item[(1)] Implausibility of small 50-60 Hz fields causing cancer \cite{AstumianEtAlRectificationPNAS1995},
\item[(2)] Plausibility of animal sensing of very small electric fields \cite{AdairEtAlTheoryDetectionWeakElectricFieldsSharkCHAOS1998},
\item[(3)] Plausibility of biological detection of small chemical reaction rates \cite{WeaverEtAl_MagneticallySensitiveChemicalReactionsBiologicalSensingSmallFieldDiff_NATURE2000}, 
and
\item[(4)] Plausibility of nsPEF causing $\sim$100 to 1000-fold more pores than conventional electroporation \cite{Schoenbach_nsPEF_Review_Bioelectromag2018}.
\end{itemize}

\noindent
\begin{figure*}[!h]
\begin{center}
\includegraphics[width=4.5in]{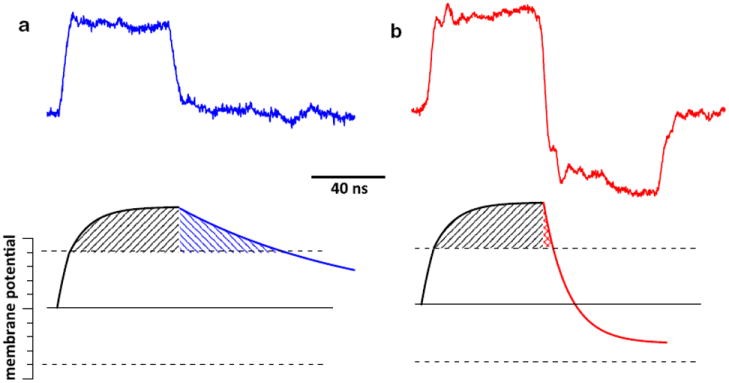}
\end{center}
\textbf{Figure 1:} ``Bipolar pulses may assist cell membrane discharge 
and reduce the membrane time above the critical voltage. 
Cell membrane is charged ({\it bottom}) 
by a monopolar pulse (\textbf{a}, {\it top}) or a bipolar pulse 
(\textbf{b}, {\it top}). The membrane 
voltage (arbitrary units) goes from the baseline ({\it solid line}) to the 
critical electroporation voltage ({\it dashed line}) and above it. The time 
when the membrane voltage exceeds the critical level is shown by {\it shading}. 
The bipolar pulse reduces this time but does
not bring the voltage below the negative critical value. 
See text for details." This figure is Figure 6 from 
\cite{PakhomovEtAlIbey_CancellationCellularResponses-nsPEF-StimulusPolarityReversal_CellMolLifeSci2014}.
\end{figure*}

Here we argue that AD (Assisted Discharge) is an implausible
mechanism of bipolar cancellation (BPC). Our critique does not 
presently apply to excitable
cells, as it considers only what is generally known about
electroporation (EP), or nanopores, in non-excitable cells. We consider
generally accepted science found in the EP literature. We purposefully do
not include experiments carried out by MURI investigators, before and
during the MURI funding. The rationale is simple and basic: We
want to understand what can be expected given established science,
mainly publications in the biophysics literature before BPC was considered. 
Accepted science does not change just because someone wants to assume that
BPC observed at $\sim$100 s is part of AD.  "Assisted discharge" means
discharge is faster (assisted) because of very large nsPEF fields that
make so many pores that the membrane conductance is greatly increased.  And 
that increased conductance is the cause of the more rapid (assisted) discharge
through the heavily porated regions of the cell membrane.
It has nothing to do with tracer molecule data and observations that occur much later, viz. at
$\sim$100 s.
Here the established scientific facts are:\\
\noindent
\begin{itemize}
\item[(1)] \underline{What is generally known about passive and porated 
cell membrane charging and discharging} 
\cite{PaulyAndSchwan_ImpedanzSuspensionKukelfoermigenTeilchenMitSchale_ZNaturforsch1959,%
StewartEtAl_CylindricalCellTransportValidationMeshing_IEEE_TransBME2005,%
Kotnik_Lightning-TriggeredElectroporation-HorizontalGeneTransferReview_PhysLifeRev2013}. See Fig. 2, 
with both passive (150 mV/cm) and heavily porated (24 kV/cm) examples.
\item[(2)] \underline{What is known broadly about nanoporation due to nsPEF}. 
Here a number of quantitative models that are broadly consistent with 
experiments are considered
\cite{DeBruinKrassowska_TheoreticalModel_SingleCellEporeI_FieldStrength_RestPotential_BPJ1999,%
DeBruinKrassowska_TheoreticalModel_SingleCellEporeII_IonicConcentrationEffects_BPJ1999,%
GowrishankarEtAl_DistributionOfFieldsPotentialsElectroporationSites_ConventionalSupraEP_CellModelExplicitOrganelles_BBRC2006,%
GowrishankarWeaver_MultiplecelluarEPsupraEP_BBRC2006,%
KrassowskaFilev_ModelingElectroporationSingleCellSphericalPoreExpansion_BPJ2006,%
SmithWeaver_ActiveMechanismsNeededSubmicrosecondMVperMeterPulses_BPJ2008,%
EsserEtAl_IntracellularManipulationConventionalEP_BPJ2010,%
LiLin_NumericalSimulationMolecularUptakeElectroporation_BChem2011,%
SozerEtAlVernier_TransportChargedSmallMoleculesElectropermeabilization-DriftDiffusion_Electroporation_BMC-Biophysics2018,%
StewartEtAl_AsympototicElectroporationTransportLaticeModel_IEEE_TransPlasmaSci2004,%
GowrishankarEtAl_NanoporeOcclusion_BPC_Mechanism_BBRC2018%
}. 
The general, established fact 
is that many small pores (nanopores) are created, which creates large membrane
conductances.
\end{itemize}

\textbf{Basic background and key definitions for BPC.} A defining feature 
of BPC is experimentally measured decrease in the 
ratio of tracer molecule influx into a cell for a bipolar pulse (BP) 
compared to a unipolar pulse (UP). Electric field pulses are followed by
cell membrane electrical discharge within a few microseconds or less. 
In sharp contrast, tracer influx
occurs over a long time (1 - 100 s) 
after the pulses. This is not trivial: after the proposed important discharge, 
the measured effect is seen 6-8 orders of magnitude displaced
in time. This means the electrically initiated molecule (tracer) transport 
is an 
exasperatingly slow process. What can account for this?! How can an early 
electrical event (discharge) lead to a greatly delayed tracer influx 
that defines BPC? This essential basic feature is missing. Further,
for both UP and BP 24 kV/cm there is immediate poration. A non-porated 
(passive) membrane is not involved, inconsistent with Fig. 6 of 
Pakhomov et al. 2014
\cite{PakhomovEtAlIbey_CancellationCellularResponses-nsPEF-StimulusPolarityReversal_CellMolLifeSci2014}.

\begin{figure*}[!h]
\begin{center}
\includegraphics[width=6.5in]{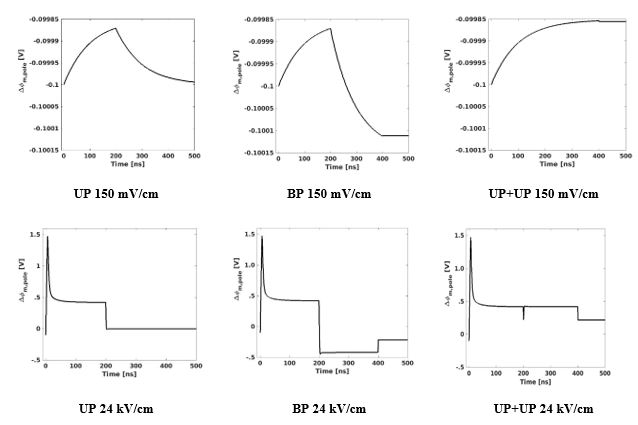}
\end{center}
\textbf{Figure 2: Response of small and large amplitude electric fields.}
These results show that assisted discharge only occurs for a passive
membrane with negligible conductance change (zero electroporation; 
low strength electric field small channel conductance).  
The 150 mV/cm, 200 ns 
unipolar pulse (top left) creates 26 pores in contrast to a 24 kV/cm,
200 ns unipolar pulse (bottom left) that creates $\mathrm{1.6\times 10^6}$
 pores in a 
5 $\mathrm{\mu m}$ radius cell. As in Fig. 6 of Pakhomov 2014  
\cite{PakhomovEtAlIbey_CancellationCellularResponses-nsPEF-StimulusPolarityReversal_CellMolLifeSci2014},
the bipolar case has a much faster discharge after the first half 
of the BP pulse. However, this occurs only for small electric fields that 
cause negligible electroporation. But to elicit BPC large fields 
($>$10 kV/cm) are used, it is a contradiction. Response of two unipolar 
pulses (with no gap) for the 150 mV/cm and 24 kV/cm fields are shown 
in the right column.  In all cases pores are created early (rapidly) 
on the first pulse and the large conduction is ``remembered".
\end{figure*}

\noindent
For perspective, passive (normal) discharge time constant of a cell is 
about 100 ns to 1 $\mathrm{\mu s}$ for typical mammalian cells
\cite{PaulyAndSchwan_ImpedanzSuspensionKukelfoermigenTeilchenMitSchale_ZNaturforsch1959,%
StewartEtAl_CylindricalCellTransportValidationMeshing_IEEE_TransBME2005,%
Kotnik_Lightning-TriggeredElectroporation-HorizontalGeneTransferReview_PhysLifeRev2013}.
Electroporated cells have spatially distributed pores of various
lifetimes, so the effective, local conductivity varies spatially 
and temporally, a complication that is readily addressed by using 
integrated cell system models
\cite{DeBruinKrassowska_TheoreticalModel_SingleCellEporeI_FieldStrength_RestPotential_BPJ1999,%
DeBruinKrassowska_TheoreticalModel_SingleCellEporeII_IonicConcentrationEffects_BPJ1999,%
GowrishankarEtAl_DistributionOfFieldsPotentialsElectroporationSites_ConventionalSupraEP_CellModelExplicitOrganelles_BBRC2006,%
GowrishankarWeaver_MultiplecelluarEPsupraEP_BBRC2006,%
KrassowskaFilev_ModelingElectroporationSingleCellSphericalPoreExpansion_BPJ2006,%
SmithWeaver_ActiveMechanismsNeededSubmicrosecondMVperMeterPulses_BPJ2008,%
EsserEtAl_IntracellularManipulationConventionalEP_BPJ2010,%
LiLin_NumericalSimulationMolecularUptakeElectroporation_BChem2011,%
StewartEtAl_AsympototicElectroporationTransportLaticeModel_IEEE_TransPlasmaSci2004,%
GowrishankarEtAl_NanoporeOcclusion_BPC_Mechanism_BBRC2018%
}.

\noindent
For the nsPEF 
(nanosecond pulsed electric fields) needed to elicit BPC there is 
an additional feature: As shown by the Schoenbach-MURI (ODU) 
\cite{Schoenbach_nsPEF_Review_Bioelectromag2018}
a broad 
finding is that nsPEF with field strengths of order 10-100 kV/cm and 
sufficiently short pulses (of order 2-1,000 ns) approximately 100-1,000 more 
pores are created, not limited to the outer, plasma membrane (PM), but 
pores are also created in many organelles within the cell, including bacterial-size 
mitochondria with double 
membranes. This means that the discharge times are much more rapid. 
All of the above is a condensed version of what is now established 
nsPEF biophysics.  

\noindent
To our knowledge the “Assisted Discharge” (AD) mechanism 
hypothesis for bipolar cancellation (BPC) was introduced by Pakhomov 
and co-workers, see their Fig. 6 and associated text, included
here as Fig. 1, but it has no equations
\cite{PakhomovEtAlIbey_CancellationCellularResponses-nsPEF-StimulusPolarityReversal_CellMolLifeSci2014}.
Subsequent publications 
cite the AD hypothesis. We find six significant flaws, identified and 
defined below.

\noindent
\underline{Flaw \#1: There is no quantitative description or prediction 
of tracer molecule influx}, the key measured biophysical quantity that 
defines BPC. 
For an illustration of what is needed for a physics-based theory or 
model we cite ``what Fermi told Dyson", a one-page commentary in Nature
\cite{Dyson_WhatFermiToldDyson-ModelingCalculating_OriginallyFromPTV_Nature2004}. 
Here Fermi states: ``One way, and this is the way I prefer, is to have a 
clear physical picture of the process that you are calculating. The
other way is to have a precise and self-consistent mathematical
formalism. You have neither." The AD hypothesis provides a qualitative
picture (Fig. 1; Fig. 6 from
\cite{PakhomovEtAlIbey_CancellationCellularResponses-nsPEF-StimulusPolarityReversal_CellMolLifeSci2014}), but there are no mathematical
calculations that quantify tracer molecule entry into a cell long after a
pulse and its associated electrical 
discharge (normal or ``assisted").

\begin{figure*}[!h]
\begin{center}
\includegraphics[width=5.5in]{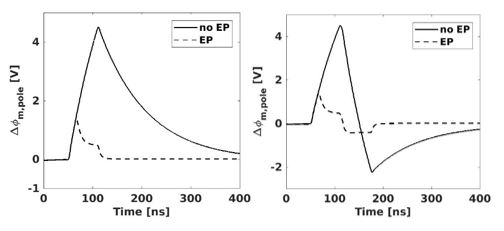}
\end{center}
\textbf{Figure 3: Passive and Standard EP model response of an isolated cell model}
in response to a 10 kV/cm, 5 ns rise/fall time, 
60 ns {\underline{unipolar pulse (left)}} and a 60 ns + 60 ns 10/-10 kV/cm 
{\underline{ bipolar pulse (right)}}. The passive model response is shown 
in solid line and 
the EP model response is shown by the dashed curve. The passive
model response for a unipolar pulse shows the transmembrane at the pole
increasing to over 4 V (for illustration only) and discharging with a 
time constant of around 100 ns. However, when EP is included, the 
response shows a reversible electrical breakdown (REB): a rapid increase
in transmembrane voltage followed by a decrease to a plateau which rapidly
reaches zero after the pulse ends ($<$1 ns). In the case 
of the bipolar pulse, the transmembrane voltage decreases much faster 
(along the lines of assisted discharge) to a smaller negative
peak before decaying to zero with a time constant of around 100 ns. With the
inclusion of EP, there is no REB peak during the second half of the BP. Instead
the transmembrane voltage remains around -0.44 V. REB is characterized by a
rapid increase in transmembrane voltage at the onset of a pulse which leads to a
burst of pore creation leading to a large increase in membrane conductance. 
This REB increase in conductance brings down the transmembrane voltage 
magnitude, limiting pore
creation even before the pulse ends. REB is seen as a plateau in transmembrane
voltage following a transient spike. As the unipolar pulse response (left) shows
that without EP, post-pulse transmembrane voltage decays with a time constant of
112 ns while significant EP causes the transmembrane voltage to decrease much
faster with a time constant of 10 ns.
\end{figure*}

\underline{Flaw \#2: There is no electroporation ``critical value", 
but instead measurement 
thresholds.}
There is no phase-transition type behavior with a rather sudden
change, here involving increasing $\mathrm{\Delta\phi_m}$
(membrane potential or transmembrane voltage). 
The horizontal dashed line(s) in Fig.~6 in 
\cite{PakhomovEtAlIbey_CancellationCellularResponses-nsPEF-StimulusPolarityReversal_CellMolLifeSci2014}
are detection or measurement thresholds that depend on both pore
creation rates at different cell membrane locations, and also on
experimental measurement capabilities. There is nothing ``critical" 
about these events. Instead they are measurement thresholds for 
either actual or Gedanken experiments, governed by signal-to-noise ratio 
(S/N). Continuum theories and molecular dynamics simulations are consistent
regarding the conditions needed to observe effects due to significant 
poration for particular experimental conditions (cell geometry and size, 
pulse waveform, etc.). Detection (measurement) theory is well known in 
physics (and therefore biophysics involving bioelectrics) \cite{AstumianEtAlRectificationPNAS1995,%
WeaverEtAl_MagneticallySensitiveChemicalReactionsBiologicalSensingSmallFieldDiff_NATURE2000,%
WeaverAstumianThermalNoiseScience1990,%
WeaverEtAl_TemperatureVariationsMolecularChangeCompetition_BPJ1999,%
VaughanAndWeaver_SignalNoiseObsv_MolecularChange_SignalNoiseGen_BEMS2005}.

\underline{Flaw \#3: Orders of magnitude different time scales.} 
For typical mammalian cells passive discharge occurs over times of 
100 nanoseconds to a few microseconds; active discharge due to applied 
fields can be much faster if the field creates many pores ({\it e.g.} nsPEF). 
But influx of tracer molecules occurs long 
afterwards, with tracer influx time scales of 1 to 100 seconds. The 
AD mechanism offers no explanation for this huge discrepancy.

\underline{Flaw \#4: Cell models must be 2D or 3D, with both an 
``inside" and ``outside"}. This feature is needed to account for 
field and ion/molecule transport both ``around" and ``through" 
the porated plasma membrane (PM). A cell model cannot be represented by 
a local planar membrane model. This is widely recognized in the 
literature
\cite{GowrishankarEtAl_NanoporeOcclusion_BPC_Mechanism_BBRC2018,%
DeBruinKrassowska_TheoreticalModel_SingleCellEporeI_FieldStrength_RestPotential_BPJ1999,%
DeBruinKrassowska_TheoreticalModel_SingleCellEporeII_IonicConcentrationEffects_BPJ1999,%
GowrishankarEtAl_DistributionOfFieldsPotentialsElectroporationSites_ConventionalSupraEP_CellModelExplicitOrganelles_BBRC2006,%
GowrishankarWeaver_MultiplecelluarEPsupraEP_BBRC2006,%
KrassowskaFilev_ModelingElectroporationSingleCellSphericalPoreExpansion_BPJ2006,%
SmithWeaver_ActiveMechanismsNeededSubmicrosecondMVperMeterPulses_BPJ2008,%
EsserEtAl_IntracellularManipulationConventionalEP_BPJ2010,%
LiLin_NumericalSimulationMolecularUptakeElectroporation_BChem2011,%
StewartEtAl_AsympototicElectroporationTransportLaticeModel_IEEE_TransPlasmaSci2004,%
GowrishankarEtAl_NanoporeOcclusion_BPC_Mechanism_BBRC2018%
}.

\underline{Flaw \#5: Small field pulses should yield Schwann (passive) 
model behavior}. Pulses too small to create nanopores should 
automatically revert to well-known passive membrane behavior 
(Fig.  2)
\cite{PaulyAndSchwan_ImpedanzSuspensionKukelfoermigenTeilchenMitSchale_ZNaturforsch1959,%
StewartEtAl_CylindricalCellTransportValidationMeshing_IEEE_TransBME2005,%
Kotnik_Lightning-TriggeredElectroporation-HorizontalGeneTransferReview_PhysLifeRev2013}.
It is not enough to explain BPC. The model must 
also be consistent with all other relevant biophysical behavior, 
e.g. response of cells to pulses too small to create large 
permeabilities or conductances in cell membranes. Passive model 
responses do not account for the many orders of
magnitude increase in membrane conductance and the resulting 
drop in transmembrane voltage magnitude (Fig. 2).

\underline{Flaw \#6: AD occurs mainly for non-porating fields 
(Figs. 2, 3)}. But, fields that elicit BPC
create over $\mathrm{10^5}$ pores, a very large number. These pores cause a 
much faster membrane discharge ($\sim$10 ns) after the pulse.

In general, the literature contains lots of results for pore behavior 
after pore creation. This undercuts the idea that one should avoid 
treating what happens above the critical membrane potential, which 
is the implication of the AD mechanism hypothesis.

\vspace{0.23in}
\Large
\textbf{Acknowledgments}
\normalsize

Supported partially by AFOSR MURI grant FA9550-15-1-0517.

\vspace{0.32in}
\textbf{References}
\vspace*{-0.52in}
\small

\def\refname{}
\bibliography{LRefEpore11-09-18mod,LRefWem09-08-18}
\bibliographystyle{unsrt}

\normalsize

\end{document}